# ARTICLE

# Hyperpolarization of Amino Acid Derivatives in Water for Biological Applications

S. Glöggler,[a] S. Wagner[b] and L.-S. Bouchard[a, c, d],

We report on the successful synthesis and hyperpolarization of N-unprotected α-amino acid ethyl acrylate esters and extensively, on an alanine derivative hyperpolarized by PHIP (4.4±1% $^{13}$C-polarization), meeting required levels for *in vivo* detection. Using water as solvent increases biocompatibility and the absence of N-protection is expected to maintain biological activity.

## Introduction

NMR is an established analytical technique broadly used in chemistry and biomedicine, recognized as a highly effective, yet insensitive, diagnostic tool. This lack of sensitivity prohibits the detection of low concentration molecules, such as metabolites or pharmaceuticals, within manageable scanning times. Various hyperpolarization techniques have been developed to overcome the sensitivity problem. These techniques rely on the creation of large non-equilibrium population differences of nuclear spin polarization in the magnetic energy sublevels of the nuclear spins. This population difference is substantially larger than the thermal (Boltzmann) population difference, potentially resulting in signal enhancements over 10,000 fold. This tremendous increase in signal offers the potential to design new contrast agents for improved disease marker detection with MRI.

Different hyperpolarization methods have been applied to pursue this goal.[1-11] For example, hyperpolarized xenon by means of spin exchange optical pumping has been successfully implemented in lung imaging and has also been caged in functionalized cryptophane structures to act as a biosensor.[1-5] Dynamic Nuclear Polarization (DNP), in which the polarization from an unpaired electron in a free radical is transferred to the nuclei, has been employed to hyperpolarize biologically active molecules and pursue their metabolism *in vivo* by monitoring the subsequent transformation of hyperpolarized products.[6-11] DNP has been the most utilized hyperpolarization to date for metabolic image; it is biocompatible and has recently been adopted in patient studies.[11]

The PHIP technique[12-13] is another very promising and more cost-effective hyperpolarization method, however, it requires the hydrogenation of a substrate molecule with an unsaturated bond using *para*-hydrogen (*p*-H$_2$), followed by a subsequent conversion of the nuclear spin singlet to a triplet state to induce polarizations.[12-13] The production of *p*-H$_2$ depends on cooling hydrogen gas to low temperatures in the presense of a catalyst. [12-13] The cooling of hydrogen gas to 25K can yield nearly 100% enrichment and thus, an almost pure quantum singlet spin order.[12-15] The application of PHIP to biomolecular imaging builds on efficient pairwise addition of *p*-H$_2$ to the substrate in aqueous solution. Based on the scarcity of discovered biological relevant substrates suited to be hyperpolarized in biologically compatible solvents, such as water, the application to living systems has lagged behind other methods. Molecular tracers must be designed and synthesized for each application of interest to incorporate viable water solubility and a hyperpolarizable PHIP moiety. The design of these molecules must provide a binding site for the *p*-H$_2$ adjacent to a heteronuclear species, such as $^{13}$C, to transfer the polarization. The heteronuclear recipient generally exhibits longer relaxation times allowing for extended *in vivo* traceability.[16]

The substrates that have been hyperpolarized to significant extents are mostly metabolites (or derivatives thereof) involved in glycolysis or the Kreb's cycle. While such applications to metabolic imaging are extremely promising, it is desirable to expand the range of molecular probes available to experimentalists and eventually for diagnostic purposes, so that additional physiological phenomenons can be explored, such as cell signaling, enzyme metabolism, and binding events to a target.

Recently, PHIP was applied to hyperpolarize select metabolites for contrast-enhanced angiography in animals.[16-20] Biologically active and biocompatible molecules that can be traced inside an organism provide the essential groundwork for molecular imaging.[21-24] A synthetic compound has also been used to detect the presense of fatty plaque in mice through chemical shift alteration induced by binding.[25] Among candidate tracer molecules, amino acids and their derivatives are of particular interest because they can potentially serve as tracers for metabolism or as indicators of binding events. The amino acids can then act as building blocks to produce peptides. One recent achievement towards the investigation of binding events includes the functionalization of a peptide, sunflower trypsin inhibitor 1, which was hyperpolarized while the enzyme remained intact.[26] Molecular interactions occur in the range of seconds to a few minutes within the decay time of the hyperpolarized signal. For example, DNP polarized glutamine is taken up and metabolized to glutamate within 20 seconds after injection *in vivo* by cancer cells and binding of cancer targeting peptides were reported *in vivo* within one to a few minutes.[27-29]

Most PHIP investigations with amino acids and their derivatives have been performed in non-aqueous solvents, such as chloroform, methanol or acetone, predominantly with protected amine functional groups. These studies[30-34] not only yielded lower polarization than reported here (in particular, a $^{13}$C polarization of 1.3% in Ref. [33]), but more importantly, the reaction exhibited little to no polarization in aqueous solution. We note that most of the previous studies were performed with 50% $p$-H$_2$ enrichment. Taking this into account, an increase in polarization by a factor of 3 can be estimated.[35] The lack of sufficient polarization however will prevent any successful clinical application. Also, to maintain biological activity and biocompatibility respectively, the design of biomarkers with active amine sites is desirable and making the use of water as solvent preferable (see discussion on amino acid polarization in Supplementary Information).

In this article, we show the synthesis of N-unprotected α-amino acid ethyl acrylate esters and in particular, an alanine derivative, for which the highest $^{13}$C polarization in D$_2$O to date was achieved when the ester was treated with *para*-hydrogen in the presence of a water-soluble rhodium complex, i.e. bis(norbornadiene) rhodium(I) tetrafluoroborate (Sigma-Aldrich). Utilizing a comparable compound, a 10% $^{13}$C-polarization with 2-hydroxyethylacrylate (HEA), and a polarization of 4.4±1% was obtained with our amino acid derivative based on alanine (nearly 100% $p$-H$_2$ enrichment). The measured polarization level can be further improved by reducing the transport time to the magnet (currently 20 s), as explained below.

HEA was chosen as the polarization-carrying moiety to obtain hyperpolarization in amino acid derivatives based on prior successes in PHIP experiments and because the hyperpolarized product, hydroxyethyl propionate (HEP) has an *in vivo* LD$_{50}$ value corresponding to that of fumarate, a metabolite naturally occurring in the citric acid cycle.[36] (For further discussion of toxic and metabolic behaviour please refer to the supplementary information.) This research represents the first significant step toward practical *in vivo* applications of PHIP techniques for peptides. The crucial contributions made by this work are three-fold: 1) generation of unprecedented $^{13}$C nuclear spin polarization of amino acid derivatives to levels sufficient for *in vivo* applications, 2) use of (deuterated) water as the solvent to increase biocompatibility; and 3) design of N-unprotected amino acid derivatives with hyperpolarizable PHIP moieties that can function as peptide building blocks to produce biologically active peptides.

## Results and Discussion

When designing a potential hyperpolarized biomarker, the critical factors are fast delivery to the target location inside the organism, as well as , sufficiently long retention of the polarization For $^{13}$C polarization spefically, the time to the target needs to be within one to two minutes. The timescale over which nuclear spins depolarize and return to their thermal equilibrium is the longitudinal relaxation time ($T_1$). As reported here, spin polarization is defined as:

$$P = (N_\beta - N_\alpha) \cdot (N_\beta + N_\alpha)^{-1},$$

in which $N_\alpha$ denotes the number of spins in the higher energy state of the nuclear magnetic energy sublevels and $N_\beta$ the number of spins in the lower energy state for a two-level system that is valid if spin $I = \frac{1}{2}$ (*e.g.*, for $^1$H and $^{13}$C nuclei). In the following, polarization will be given as a percentage $P \cdot 100\%$. Essentially, a tracer that achieves both high polarization and a long $T_1$ is what we are aiming for.

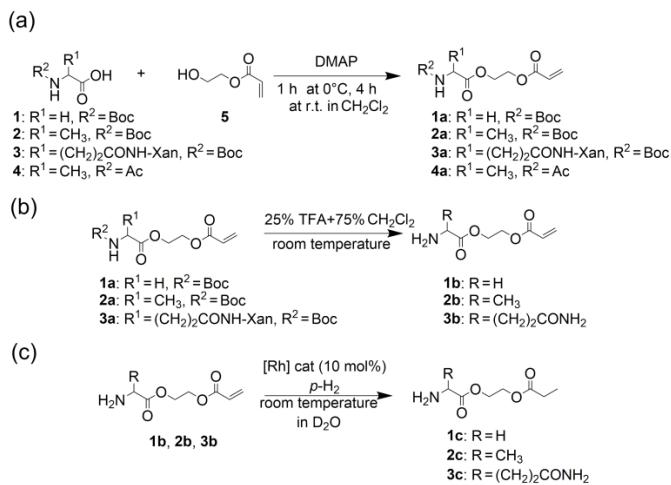

**Figure 1.** (a) Synthesis of the protected amino acid ethyl acrylate esters from N-protected amino acids and hydroxyethylacrylate (b) Deprotection of the amine function with TFA yields N-unprotected α-amino acid derivatives. (c) Hydrogenation with *para*-hydrogen in deuterated water catalyzed by a water-soluble rhodium complex.

HEA has been frequently used in PHIP hyperpolarization studies, including at times as a MRI contrast agent for angiography in animals.[16-20-31] It is commercially available in both, deuterated- and $^{13}$C-labelled forms. When partly deuterated, one of the $^{13}$C nuclei of the acrylate functionality possesses a $T_1$ close to two minutes in deuterated solvents in low magnetic fields, and a polarization of $P > 10\%$ is routinely detected in the magnet through the use of a custom-built polarizer apparatus.[16,37] The α-amino acids derivatized in this study were chosen to be glycine, alanine and glutamine for their potential relevance as tracers in molecular imaging experiments. Glycine is the simplest proteinogenic amino acid and has recently been identified as a metabolite associated with cancer proliferation, it is used for cellular purine synthesis, a building block for nucleic acids and thus, DNA.[38] The metabolic pathways of alanine are closely connected to pyruvate, a metabolite often overexpressed in cancer cells.[9] Glutamine is currently being discussed as a potential tracer in MRI studies as it is overexpressed in some cancer cells.[39] Multiple studies with positron emission tomography tracers or amino acid modified gadolinium complexes indicated that significantly altered amino acids, even with considerably larger modifications than presented here, are still biologically active and internalized by cells according to the same pathways as amino acids on a minute timescale. This has been evaluated for alanine and glutamine derivatives.[39-42]

The amino acid derivatives used in our experiments (Figure 1) were synthesized, as detailed in the Methods section. The hyperpolarized products yielded amino acid ethyl propionate esters (**1c–3c**; Figure 1c) that show a characteristic triplet and quartet at 1.0 and 2.4 ppm, respectively, in the $^1$H NMR spectrum for the investigated substrates (Figure 2). Typical yields for the proton experiments ranged from 5 to 10% for unoptimized mixing conditions. The reported polarization values have been normalized to the yield. The observed phase pattern corresponds to $I_z$-$S_z$ magnetization as expected from an ALTADENA experiment but is shown in magnitude mode for comparison with the thermal spectrum.[43] Hydrogenation experiments were conducted at two different temperatures: 20 °C and 80 °C.

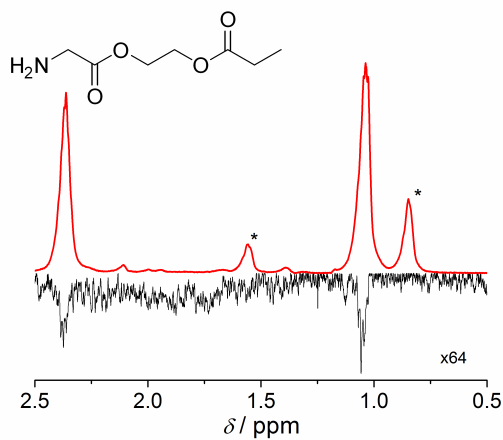

**Figure 2.** Hydrogenation of glycine derivative in D$_2$O using water-soluble rhodium complex. Top trace: Hyperpolarized section of the proton spectrum following hydrogenation performed with $p$-H$_2$ at 80 °C. After the esters were treated with $p$-H$_2$, physical transport of the sample to the NMR magnet took 10 s. Bottom trace: Thermal polarization of propionate protons following equilibration at 14.1 T and 20 °C. The bottom trace has been magnified by a factor of 64 and both spectra are shown in magnitude (absolute value) mode. Peaks attributed to the decomposition products of the catalyst are marked with an asterisk. The level of polarization in the hyperpolarized product corresponds to 0.73 % after transport.

Notably, the proton polarization of all amino acid derivatives was at least a factor of 2 higher at the elevated temperature (Table 1). Hyperpolarization of the glycine and alanine derivatives, both containing a single amine function, resulted in a measured proton polarization of $P$(gly)=0.73% and $P$(ala)=0.70 %. All experiments were performed at a pH = 6.5 ± 0.5 as indicated by pH test paper. A large difference from a neutral pH may result in a better polarization but has a disadvantage: the polarization product requires dilution in buffer prior to injection to regulate the pH value to a biocompatible level. Dilution of the compounds can result in concentrations of the potential tracers that are too low. Table 1 summarizes the determined proton polarization values of all hyperpolarized molecules investigated. The $T_1$ relaxation times of the protons range between 2–6 s (see Supplementary Information) at the detection field. However, due to the transportation time between polarization and detection (10 s), significant polarization is lost before acquisition. Since $T_1$ is not constant during the transportation process from earth's magnetic field into high magnetic field, the loss in polarization cannot be precisely quantified here but is considered to be substantial on account of validation through $^{13}$C transfer experiments. An example $^1$H NMR spectrum of the hyperpolarized region of the glycine derivative is shown in Figure 2. Here, a signal enhancement of ≈200 relative to the thermal polarized signal at B$_0$ = 14.1 T (600 MHz proton frequency) can be clearly discerned. In high temperature experiments, additional hyperpolarized peaks can be identified, which are attributed to the decomposition of the catalyst. These peaks are discussed in more detail in the Supplementary Information.

**Table 1.** Proton polarization of the hyperpolarized amino acid derivatives

| Hyperpolarized species | | Polarization at 20 °C [%] | Polarization at 80 °C [%] |
|---|---|---|---|
| 1c | H$_2$N-CH$_2$-C(O)O-CH$_2$CH$_2$-O-C(O)-CH$_2$CH$_3$ | 0.33 | 0.73 |
| 2c | H$_2$N-CH(CH$_3$)-C(O)O-CH$_2$CH$_2$-O-C(O)-CH$_2$CH$_3$ | 0.13 | 0.70 |
| 3c | glutamine derivative structure | 0.02 | 0.52 |
| 4b | acetyl-protected alanine derivative structure | 0.48 | 1.02 |
| 5a | HO-CH$_2$CH$_2$-O-C(O)-CH$_2$CH$_3$ | 0.74 | 1.26 |

For the hyperpolarized glutamine derivative, which contains two unprotected amine groups, proton polarization of only 0.02 % was observed at 20°C, but 0.52% was observed at 80°C. Previous studies on rhodium transition metal complexes with phosphine ligands indicate that catalytic activity may be reduced if unprotected amine groups are present in the reaction mixture during hydrogenation; the observed trend in polarization can be deduced as follows: The reaction proceeds faster with increasing temperature and the quantum correlation of *para*-hydrogen used to create hyperpolarization can be transferred more effectively. This effect is contrary to the progressive deactivation of the catalyst by free amine groups, which leads to catalyst decomposition slowing down the addition of *para*-hydrogen and leading to a decrease in observable proton polarization.[44] In this case, the catalyst is sacrificed in the reaction to produce more product carrying a higher polarization.

To verify the effect of the free amine on the achieved polarization, a protected alanine-derived amine was synthesized. Because the commonly used *tert*-butoxycarbonyl (Boc)-protected amino acid ethyl acrylate ester is sparingly soluble in water, an acetyl protecting group was selected to protect the amine (**4a**). Hyperpolarization was performed accordingly to the procedure described with 10 mol % catalyst and 5 bar *para*-hydrogen pressure. In the experiments, the level of proton polarization achieved of the acetyl protected alanine derivative was approximately a factor of four ($P$ = 0.48 %) higher at 20 °C and ≈ 1.5 ($P$ = 1.02 %) at 80 °C relative to that obtained with the non-acetylated form, which suggests that the free amine group deactivates the rhodium complex, probably as described above. With amine protection, the polarization is within 80% of the hydroxyethyl propionate standard (HEP, **5a**).

Subsequently, transfer of the proton polarization to a $^{13}$C nucleus was investigated using the alanine derivative. When HEP is polarized with a device that features optimal reaction conditions and mixing techniques to react *para*-hydrogen with the substrate and pulse sequences for polarization transfer, a $^{13}$C polarization of more than 10 % can be reached.[16,45] By following the same approach, $^{13}$C and deuterium-labelled HEA were utilized for the synthesis of a labelled alanine derivative and hyperpolarized, followed by subsequent detection in a 9.4 T NMR magnet (Figure 3). The polarizer used in this experiment has been recently described.[46] The $T_1$ values of $^{13}$C nuclei in carbonyl-HEP moieties in non-deuterated molecules and in deuterated solvent are 14–17 s (with air present at 11.7 T, see the Supplementary Information for detailed $T_1$ data) and 21.5 s in the deuterated compound at 14.1 T. These values not only affect the polarization transfer but are typical values necessary for *in vivo* studies; however, they are expected to be smaller in non-deuterated solvents and *in vivo*.[47] (As mentioned previously, a minimum of tens of seconds to minutes are required for transport of tracer molecules to targets of interest.) However, it has been pointed out that the relaxation times of $^{13}C$ in carbonyl groups are strongly influenced by the chemical shift anisotropy, which is proportional to the magnetic field strength.[48] The determined $T_1$ = 21.5 s was measured at 14.1 T. Close to the earth's magnetic field, where polarization transfer and transport take place, $T_1$ is expected to be much longer and may be close to 2 minutes as for HEA. The yielded polarization was $P$ = 4.4±1 % for the alanine derivative and $P$ = 10 % for HEA with 100% conversion to the desired product. These values were obtained in a 9.4 T field after an unoptimized transportation process that took 20 s. The 20 s time was due to the physical transport of the sample tube from the polarizer to the center of the magnet. Improvements should be possible, shortening the distance from the polarizer to the magnet. Extrapolating back to time $t$ = 0, which would be prior to the transportation process, (to compare the polarization e.g. with Ref. [49]) the $^{13}C$ polarization of the alanine derivative immediately following chemical reaction is estimated to be $P$ = 12 %, if a $T_1$ value of 21.5s is assumed. This gives an indication of the maximally achievable $^{13}C$ polarization value but may in effect be smaller due to the $T_1$ being greater in low magnetic fields during the transport. Hence, the $^{13}C$ polarization level for this amino acid derivative may be comparative to polarization levels achieved with metabolites such as pyruvate.[11] This result highlights that, despite the presence of free amine groups, the polarization required for *in vivo* applications is feasible. A $^{13}C$ polarization of $P$ = 12 % would correspond to a signal enhancement of ≈ 86,000 compared to thermal polarizations in a standard clinical MRI scanner ($B_0$ = 1.5 T) or ≈ 17,000,000 at 5 mT (see the field strength of interest in Ref. [49]). Key to achieving such high levels of polarization for amino acid derivatives is the use of the sequence introduced by Goldman et al.[16] This sequence works perfectly for the demonstrated derivatives but may not always be applicable for some amino acid derivatives, as discussed in the supplementary information.

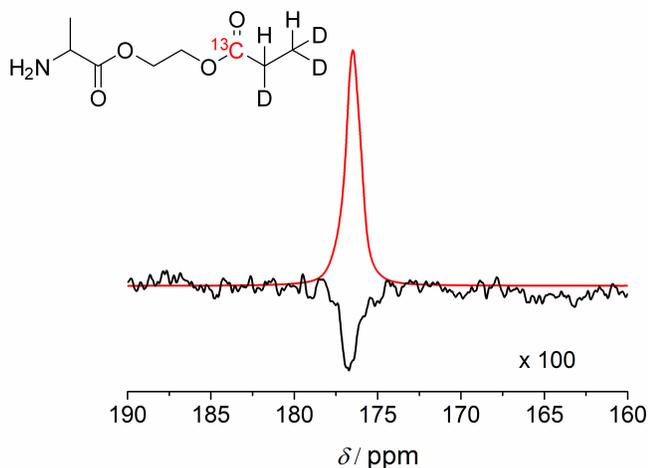

**Figure 3.** The substrate, deuterium-labeled **2c**, was hyperpolarized using a PHIP polarizer device followed by a polarization transfer sequence to $^{13}C$ nuclei and physical transport of the polarized substrate to the NMR magnet for measurement. Top trace: Hyperpolarized signal from $^{13}C$ NMR spectroscopy and deuterium-labeled (real spectrum) **2c**. The physical transport to the NMR magnet took 20s. Bottom trace: Thermally polarized signal after 16 scans following equilibration at 9.4T and 20°C. This spectrum was magnified 100 times. The level of polarization in the hyperpolarized product corresponds to 4.4±1 % after transport.

## Conclusions

The N-unprotected α-amino acid ethylacrylate esters based on glycine, alanine and glutamine with hyperpolarizable PHIP moieties were successfully synthesized and applied. The alanine derivative yielded $^{13}C$ polarization levels in sufficient concentrations (10 mM) in deuterated water and sufficient signal for *in vivo* applications.[9,19] The high polarization, which correlates with the decrease of free amine moieties in the synthesized molecules, indicates a gain of catalyst activity during the hydrogenation reaction. Our findings suggest that amino acid based acrylates can potentially serve as new contrast agents either by themselves or as labelled "building blocks" within peptides that retain partial bioactivity with a limited number of unprotected amine groups. Although, amino acid derivatives are likely to be taken up by cells, it is unclear whether the derivatives take part in metabolic processes that can be observed. Work is currently underway in our laboratory to investigate the metabolic behavior and to extend this method to small cancer-binding peptides that may be of particular interest in combination with recently developed nanoparticles showing significant level of polarization utilizing heterogeneous PHIP in water, thereby mitigating catalyst toxicity effects.[50]

## Acknowledgements

The authors gratefully acknowledge financial support from NSF, grant CHE-1153159, equipment grant CHE-1048804, the Jonsson Comprehensive Cancer Center (JCCC) at UCLA and the Arnold and Mabel Beckman Foundation through a Young Investigator Award. Manuscript editing by Schlicht Scientific Editing. The authors thank Philipp Schleker for constructive discussions.

## Notes and references

[a] Department of Chemistry and Biochemistry, University of California at Los Angeles, Los Angeles, California 90095-1569, USA.
[b] Biomedical Imaging Research Institute, Cedars Sinai Medical Center, 8700 Beverly Blvd, Davis Building G149E, Los Angeles, California 90048-1804, USA.
[c] California NanoSystems Institute, 570 Westwood Plaza, Building 114, Los Angeles, California 90095-1569, USA.
[d] Department of Bioengineering, University of California at Los Angeles, 420 Westwood Plaza, RM 5121 Engineering V, P.O. Box 951600, Los Angeles, California 90095-1569, USA.

Electronic Supplementary Information (ESI) available: [synthesis, para-hyorgen experiments, relaxation times of amino acid derivatives, catalyst decomposition and remarks on amino acid PHIP]. See DOI: 10.1039/c000000x/

# Hyperpolarization of Amino Acid Derivatives in Water for Biological Applications

Stefan Glöggler, Shawn Wagner, Louis-S. Bouchard

# Supporting Information

**Contents**

# S1 Materials and Methods

**Chemicals**

N,N'-Dimethylformamide (DMF), diethyl ether methylene chloride (DCM), sodium bicarbonate (NaHCO$_3$) and concentrated HCl (10 M) were purchased from Thermo Fisher Scientific. Absolute methanol under inert gas, pentane, trifluoroacetic acid (TFA), Dicyclohexylcarbodiimide (DCC), dimethylaminopyridine (DMAP), Boc-Gly-OH (**1**), Boc-Ala-OH (**2**), Ac-Ala-OH (**4**), 2-hydroxyethyl acrylate (**5**), bis(norbornadiene) rhodium(I) tetrafluoroborate (**6**) and 1,4-Bis[(phenyl-3-propanesulfonate) phosphine] butane disodium salt (**7**) were purchased from Sigma-Aldrich. Boc-Gln-Xan-OH (**3**) was purchased from Bachem and CDCl$_3$ and D$_2$O (99.8%) from Cambridge isotopes. All chemicals were used as received without further purification.

**NMR Spectroscopy**

NMR spectra were recorded on either a Bruker AV600, a Bruker DRX500 or Bruker AV500 with cryoprobe. All experiments were performed at room temperature and the chemical shift data are reported in ppm for $^1$H and $^{13}$C spectra relative to tetramethylsilane or 4,4-dimethyl-4-silapentane-1-sulfonic acid.

**Mass Spectrometry**

Mass Spectrometry was performed on a LCT Premier XE of Micromass MS Technologies by electrospray ionization. Masses are given in m/z and were compared to the exact calculated mass.

## S2 Synthetic procedures

**Synthesis of 1a**

In 15 mL methylene chloride, 1.0 mmol (175.2 mg) Boc-Gly-OH and 0.25 mmol (30 mg) DMAP were dissolved and 1.0 mmol (116.1 mg, 114.8 µL) 2-hydroxyethyl acrylate added. The solutions was cooled to 0°C and 1.1 mmol (227.0 mg) dicyclohexylcarbodiimide in 5 mL methylene chloride added. It was stirred at 0°C for 1 hour, was subsequently allowed to warm up to room temperature and stirred for an additional 4 hours. Afterwards, the solution was concentrated to dryness, the residue taken up in 15 mL methylene chloride, filtered and washed two times with 5 mL HCl in water (0.5 M) and two times with a saturated aqueous $NaHCO_3$ solution. The organic phase was collected, residual solvent removed and the product further purified by column chromatography with 95% DCM and 5% MeOH. The product was received as colorless oil. $^1$H NMR (600 MHz, [D]CDCl$_3$, 25°C): $\delta$=6.32 (d, $^3J$(H,H)=17.4 Hz, 1H; CH), 6.07 (dd, $^3J$(H,H)=17.4 Hz, $^3J$(H,H)=10.7 Hz, 1H; CH), 5.87 (d, $^3J$(H,H)=10.7 Hz, 1H; CH), 4.37 (m, 4H; CH$_2$,CH$_2$), 3.93 (2d, $^2J$(H,H)=9.0 Hz, 2H; CH$_2$), 1.43 ppm (s, 9H, 3CH$_3$); $^{13}$C NMR (151 MHz, [D]CDCl$_3$, 25°C): $\delta$=170.45, 165.74, 155.58, 134.47,127.72, 79.95, 62.80, 61.87, 42.19, 28.16 ppm; MS: m/z (%):calculated: 296.1110, found: 296.1108 [M+Na]$^+$.

**Synthesis of 2a**

Synthesis of **2a** was achieved according to the same procedure as described for **1a** by using **2** as a reactant instead of **1**. The product was received as colorless oil. $^1$H NMR (600 MHz, [D]CDCl$_3$, 25 °C): $\delta$=6.32 (d, $^3J$(H,H)=17.4 Hz, 1H; CH), 6.07 (dd, $^3J$(H,H)=17.4 Hz, $^3J$(H,H)=10.7 Hz, 1H; CH), 5.87 (d, $^3J$(H,H)=10.7 Hz, 1H; CH), 4.37 (m+q, 4H+1H; CH$_2$,CH$_2$,CH), 1.43 (s, 9H, 3CH$_3$), 1.37 ppm (d, $^3J$(H,H)=7.4 Hz, 3H, CH$_3$); $^{13}$C NMR (151 MHz, [D]CDCl$_3$, 25°C): $\delta$=173.08, 165.70, 154.96, 131.40,127.75, 79.77, 62.70, 61.89, 49.04, 28.17, 18.43 ppm; MS: m/z (%):calculated: 310.1266, found: 310.1275 [M+Na]$^+$.

**Synthesis of 3a**

0.5 mmol (mg) **3** and 0.25 mmol (30 mg) DMAP were dissolved in 0.5 mL DMF, followed by addition of 0.5 mmol (58.1 mg, 57.4 µL) 2-hydroxyethyl acrylate and 15 mL DCM. The solutions was cooled to 0°C and 1.1 mmol (227.0 mg) dicyclohexylcarbodiimide in 5 mL methylene chloride added. It was stirred at 0°C for 1 hour, was subsequently allowed to warm up to room temperature and stirred for an additional 5 hours. Afterwards, the solution was concentrated to dryness, the residue taken up in 15 mL methylene chloride, filtered and washed two times with 5 mL HCl in water (0.5 M) and two times with a saturated aqueous $NaHCO_3$ solution. The organic phase was collected, residual solvent removed, washed with pentane and the product further purified by column chromatography with 95% DCM and 5% MeOH. The product was received as a colorless solid. $^1$H NMR (500 MHz, [D]CDCl$_3$, 25°C): $\delta$= 7.47 (m, 2H, CH), 7.30 (m, 2H, CH), 7.11 (m, 4H, CH), 6.52 (m, 1H, CH), 6.34 (d, $^3J$(H,H)=17.4 Hz, 1H; CH), 6.03 (dd, $^3J$(H,H)=17.4 Hz, $^3J$(H,H)=10.7 Hz, 1H; CH), 5.78 (d, $^3J$(H,H)=10.7 Hz, 1H; CH), 4.35 (m+t, 4H+1H; CH$_2$,CH$_2$,CH), 2.29 (m, 2H, CH$_2$), 2.21 (m, 1H, CH), 2.04 (m, 1H, CH), 1.41 ppm (s, 9H, 3CH$_3$); $^{13}$C NMR (126 MHz, [D]CDCl$_3$, 25°C): $\delta$=171.99, 170.68, 165.73, 155.40, 151.03, 131.59, 129.51, 129.15, 127.58, 123.48, 120.95, 116.52, 80.04, 63.05, 61.87, 52.91, 43.84, 32.34, 29.60, 28.26 ppm; MS: m/z (%):calculated: 547.2056, found: 547.2036 [M+Na]$^+$.

**Synthesis of 4a**

Synthesis of **4a** was achieved according to the same procedure as described for **3a** by using **4** as a reactant instead of **3**. The product was received as colorless oil. . $^1$H NMR (500 MHz, [D$_2$]D$_2$O, 25°C): $\delta$=6.32 (d, $^3J$(H,H)=17.4 Hz, 1H; CH), 6.07 (dd, $^3J$(H,H)=17.4 Hz, $^3J$(H,H)=10.7 Hz, 1H; CH), 5.87 (d, $^3J$(H,H)=10.7 Hz, 1H; CH), 4.32 (m, 4H; CH$_2$,CH$_2$), 4.27 (q, $^3J$(H,H)=7.4 Hz, 1H, CH), 1.87 (s, 3H, CH$_3$), 1.26 ppm (d, $^3J$(H,H)=7.4 Hz, 3H, CH$_3$); $^{13}$C NMR (126 MHz, [D$_2$]D$_2$O, 25°C): $\delta$=174.38, 173.38, 167.99, 132.80, 127.07, 63.27, 62.46, 48.66, 21.28, 15.64 ppm; MS: m/z (%):calculated: 252.0848, found: 252.0852 [M+Na]$^+$.

**Amine-deprotection of 1a and 2a**

The Boc-group of **1a** and **2a** was removed by dissolving them in excess of a 1:1 by volume mixture of DCM and TFA and subsequent stirring for 30 minutes at room temperature. The residues were taken up in deionized water, filtered and the solvent removed under vacuum to yield product **1b** and **2b** as colorless oils. All deprotected derivatives were found to degrade rapidly and were stored under inert gas at 4°C. **1b**: $^1$H NMR (500 MHz, [D$_2$]D$_2$O, 25°C): $\delta$=6.32 (d, $^3J$(H,H)=17.4 Hz, 1H; CH), 6.07 (dd, $^3J$(H,H)=17.4 Hz, $^3J$(H,H)=10.7 Hz, 1H; CH), 5.87 (d, $^3J$(H,H)=10.7 Hz, 1H; CH), 4.42 (m, 4H; CH$_2$,CH$_2$), 3.82 ppm (s, 2H; CH$_2$); $^{13}$C NMR (126 MHz, [D$_2$]D$_2$O, 25°C): $\delta$=168.05, 167.80, 132.67, 127.00, 64.10, 62.34, 39.85 ppm; MS: m/z (%):calculated: 174.0766, found: 174.0762 [M+H]$^+$. **2b**: $^1$H NMR (500 MHz, [D$_2$]D$_2$O, 25°C): $\delta$=6.32 (d, $^3J$(H,H)=17.4 Hz, 1H; CH), 6.07 (dd, $^3J$(H,H)=17.4 Hz, $^3J$(H,H)=10.7 Hz, 1H; CH), 5.87 (d, $^3J$(H,H)=10.7 Hz, 1H; CH), 4.40 (m, 4H; CH$_2$,CH$_2$), 4.09 (q, $^3J$(H,H)=7.4, 1H, CH); 1.43 ppm (d, $^3J$(H,H)=7.4, 3H; CH$_3$); $^{13}$C NMR (126 MHz, [D$_2$]D$_2$O, 25°C): $\delta$=170.36, 168.02, 132.70, 127.00, 64.21, 62.32, 48.56, 14.87 ppm; MS: m/z (%):calculated: 188.0923, found: 188.0915 [M+H]$^+$.

**Amine-deprotection of 3a**

3a was dissolved in an excess of a 1:1: by volume mixture of DCM and TFA. After 5 minutes of stirring at room temperature the solvents were removed under vacuum.

Diethyl ether was added (2mL) twice and decanted. The remaining residue was washed three times with pentane and yielded 3b as colorless wax. $^1$H NMR (500 MHz, [D$_2$]D$_2$O, 25°C): $\delta$=6.32 (d, $^3J$(H,H)=17.4 Hz, 1H; CH), 6.07 (dd, $^3J$(H,H)=17.4 Hz, $^3J$(H,H)=10.7 Hz, 1H; CH), 5.87 (d, $^3J$(H,H)=10.7 Hz, 1H; CH), 4.40 (m, 4H; CH$_2$,CH$_2$), 4.08 (t, $^3J$(H,H)=7.0, 1H, CH), 2.36 (m, 2H, CH$_2$), 2.09 ppm (m, 2H, CH$_2$); $^{13}$C NMR (126 MHz, [D$_2$]D$_2$O, 25°C): $\delta$=176.57, 169.40, 167.96, 132.74, 126.97, 64.32, 62.24, 51.99, 30.08, 25.15; MS: m/z (%):calculated: 245.1102, found: 245.1105 [M+H]$^+$.

**Synthesis of the water soluble PHIP catalyst**

1,4-bis-[(phenyl-3-propane sulfonate) phosphine]butane (norbornadiene) rhodium(I)tetrafluoroborate (**8**), which has previously been used as water soluble PHIP catalyst has been synthesized in a different procedure as reported:[S1] Under argon atmosphere, in two different schlenk flasks, 10 µmol (7.5 mg) of **6** and 10 µmol (11.3 mg) of **7** were dissolved in 5 mL absolute methanol. After stirring for 30 minutes, **7** was slowly added over 1 minute to **6** and stirring continued for 1 hour followed by concentration to dryness under vacuum.

## S3 NMR spectra

NMR spectra are shown of the reactants (**1b-3b**, **4a**), which have been used for the hyperpolarization experiments.

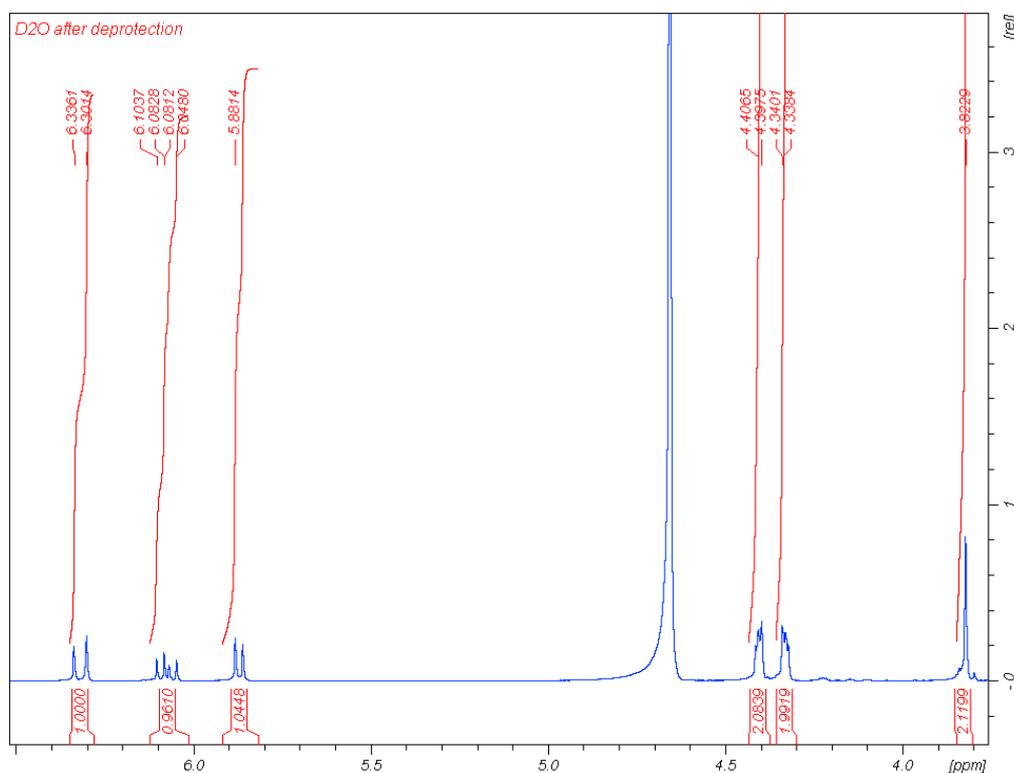

**Fig. S1**. $^1$H NMR spectrum of **1b** at $B_0$ = 11.7 T (500 MHz $^1$H frequency) acquired with 8 scans.

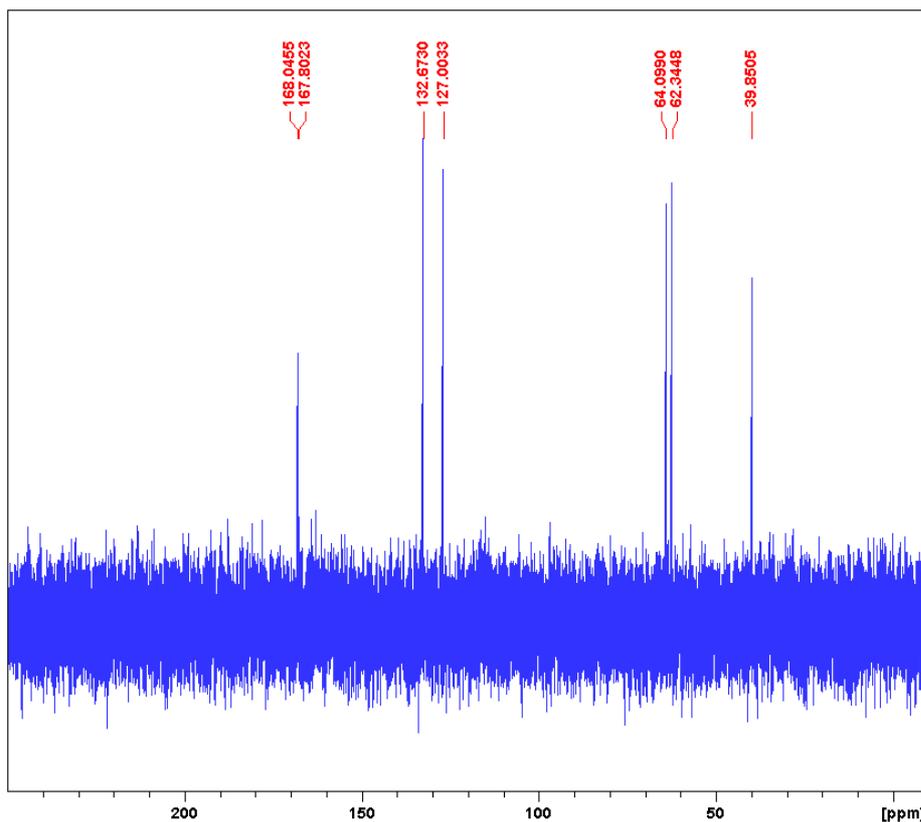

**Fig. S2**. $^{13}$C NMR spectrum of **1b** at $B_0$ = 11.7 T (126 MHz $^{13}$C frequency) acquired with 64 scans.

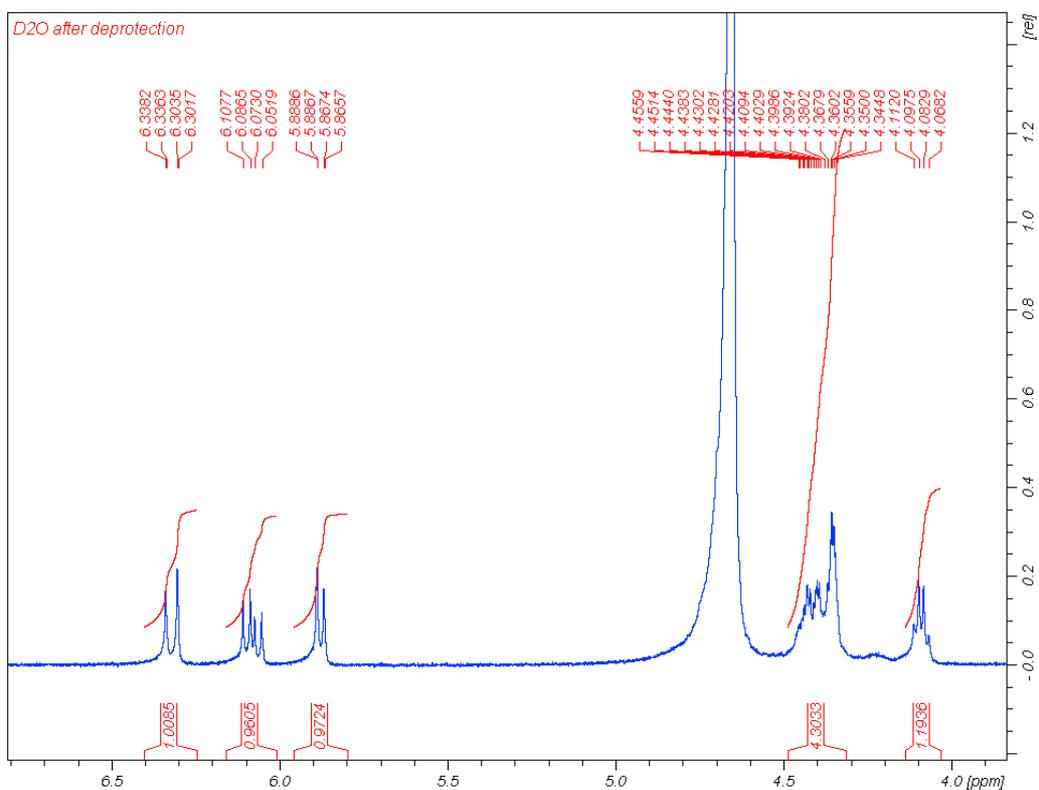

**Fig. S3**. $^1$H NMR spectrum of **2b** at $B_0$ = 11.7 T (500 MHz $^1$H frequency) acquired in a single scan.

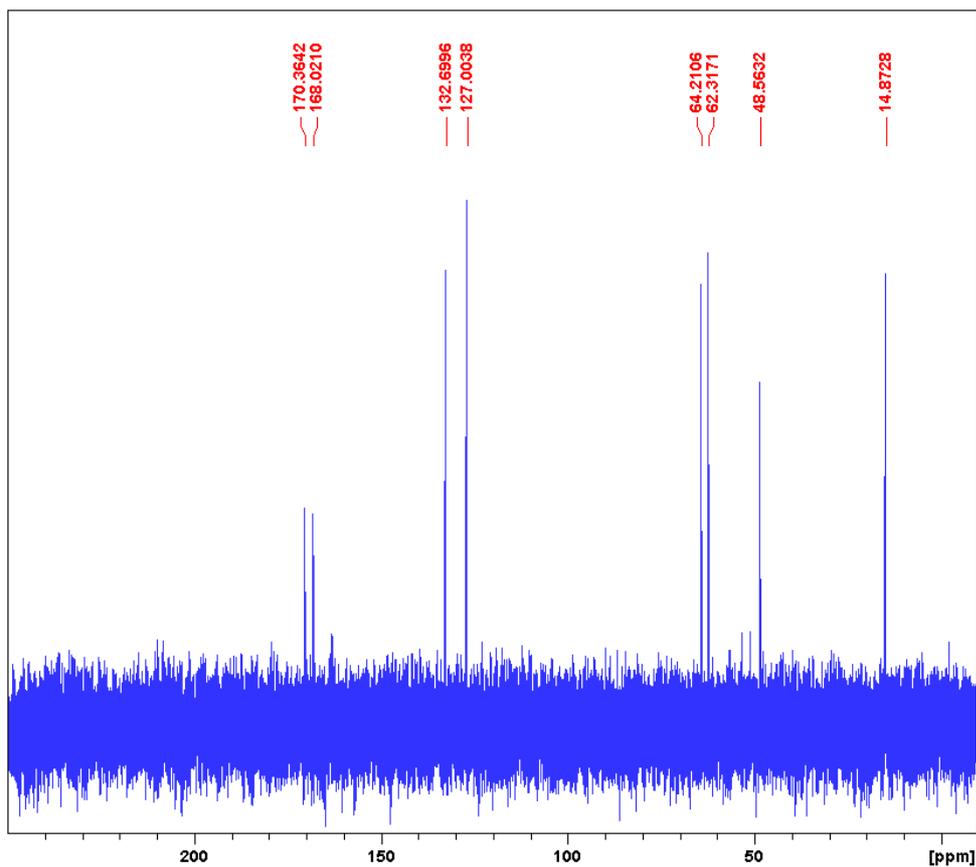

**Fig. S4**. $^{13}$C NMR spectrum of **2b** at $B_0 = 11.7$ T (126 MHz $^{13}$C frequency) acquired with 128 scans

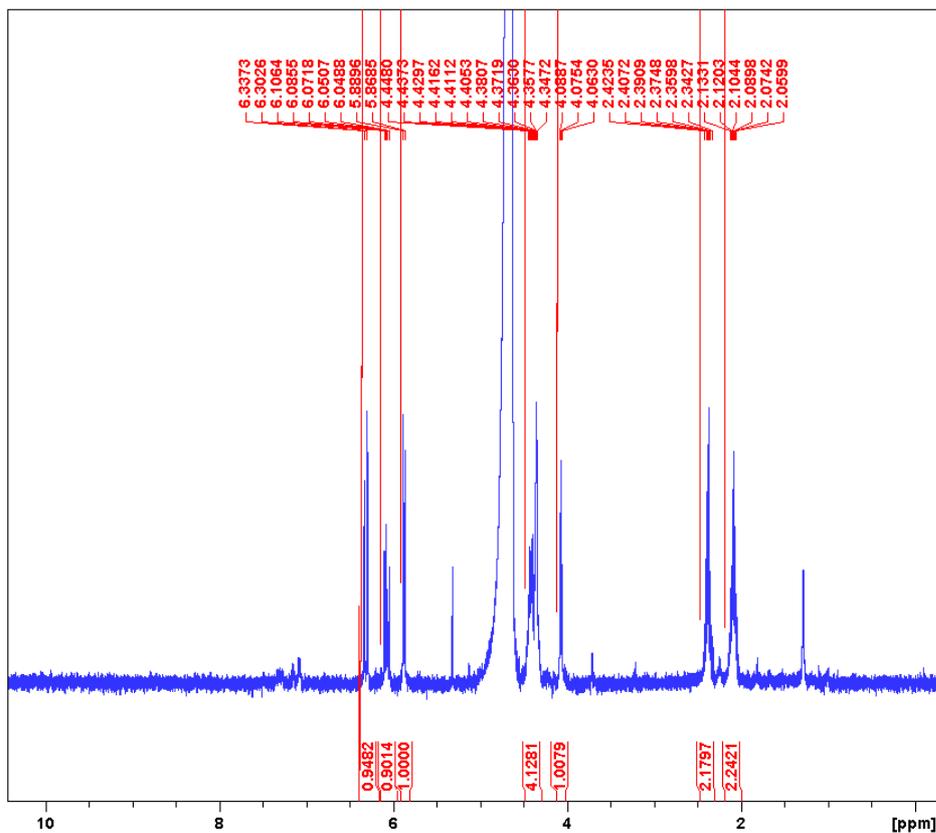

**Fig. S5**. $^{1}$H NMR spectrum of **3b** at $B_0 = 11.7$ T (500 MHz $^{1}$H frequency) acquired with 8 scans

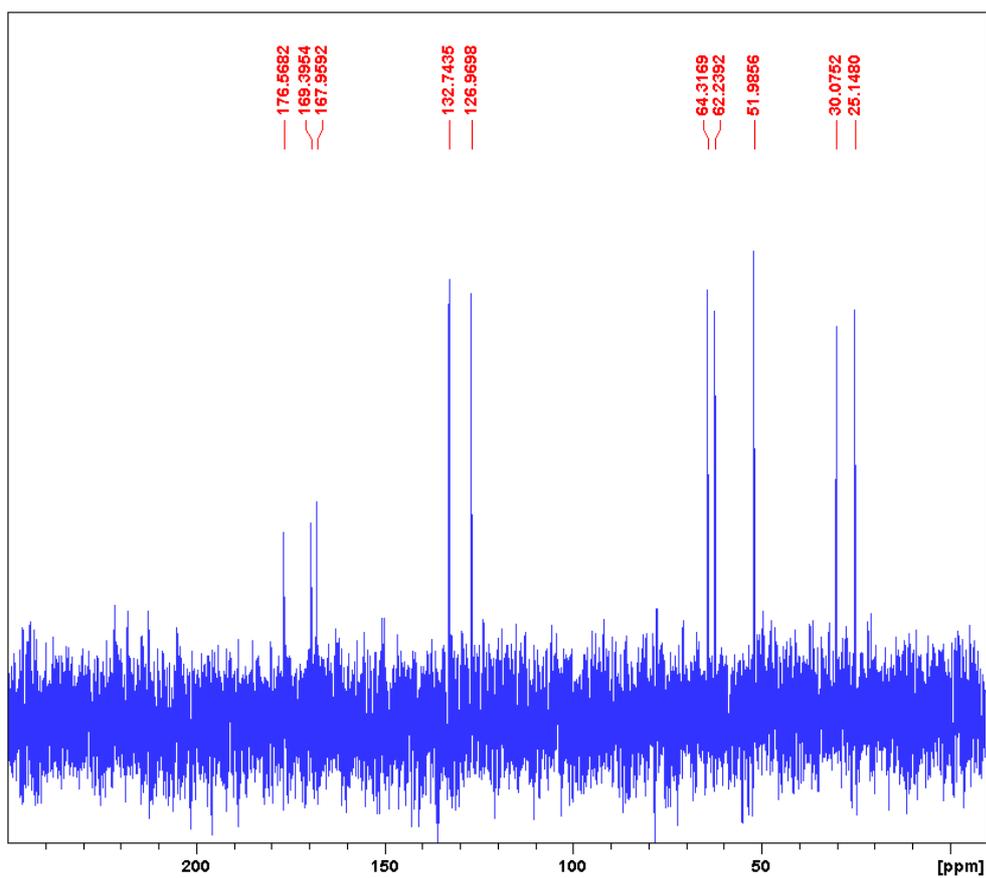

**Fig. S6**. $^{13}$C NMR spectrum of **3b** at $B_0$ = 11.7 T (126 MHz $^{13}$C frequency) acquired with 512 scans

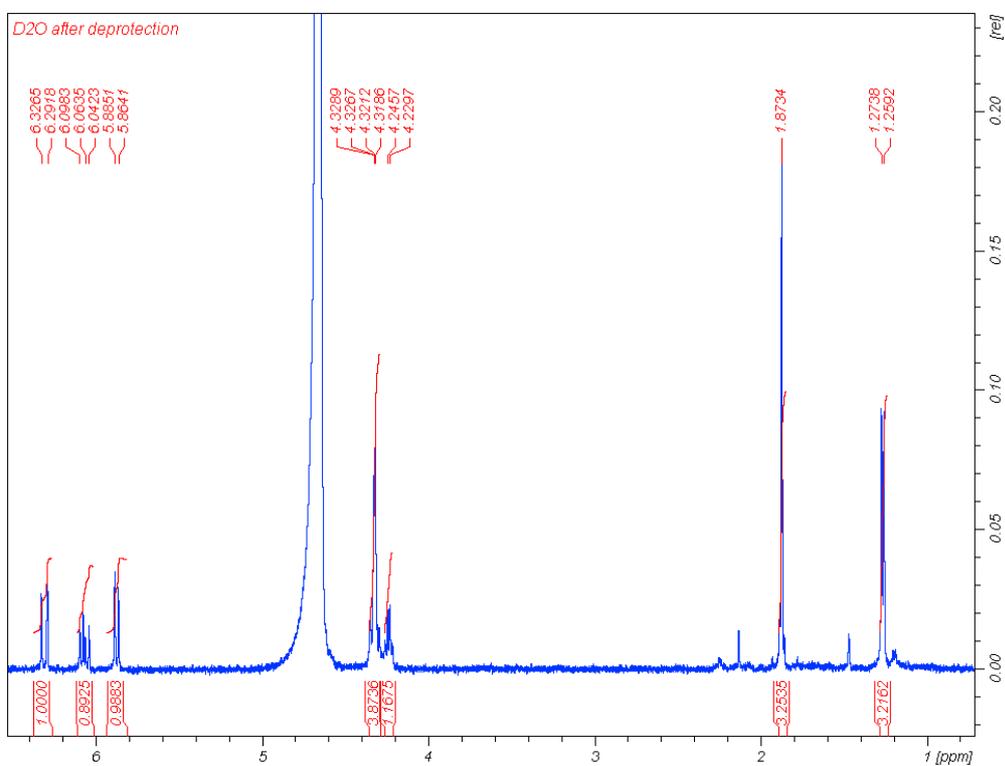

**Fig. S7**. $^1$H NMR spectrum of **4a** at $B_0$ = 11.7 T (500 MHz $^1$H frequency) acquired with 8 scans.

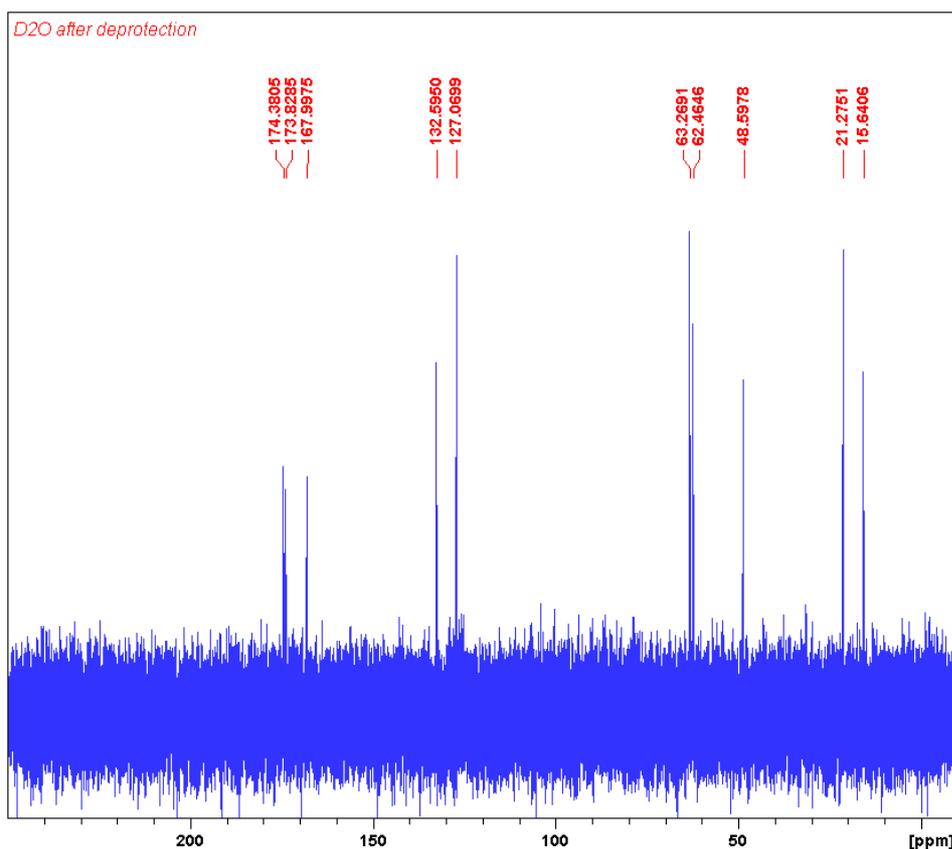

**Fig. S8**. $^{13}$C NMR spectrum of **4a** at $B_0$ = 11.7 T (126 MHz $^{13}$C frequency) acquired with 512 scans.

## S4 *Para*-hydrogen experiments

Proton *para*-hydrogen experiments were performed on a Bruker AV600 spectrometer ($B_0$ = 14.1 T). *Para*-hydrogen of about 95% *para*-state enrichment was produced using a commercial polarizer located at Cedars Sinai Medical Center.[S2] Samples were prepared in 5 mm J. Young tubes from New Era under inert gas with sample (**1b-3b**, **4a**) concentrations of 25 mM and 2.5 mM catalyst concentration in 0.5 mL $D_2O$. Each sample was pressurized with 5 bars of *para*-hydrogen, shaken for 10 s in the earth's magnetic field (ALTADENA conditions) and transported to the center of the magnet within 10 s, where the spectrum was recorded in a single scan (45°-pulse). The experiments were repeated three times with different samples. After the hyperpolarization experiment, a spectrum was recorded with the formed product in thermal equilibrium, the signal enhancement and the corresponding nuclear magnetic polarization calculated. All experiments were conducted at pH = 6.5±0.5. Polarization transfer experiments were performed on a home-built polarizer[S2] with a 10 mM solution of the unprotected alanine derivative and 2.2 mM catalyst concentration. The sample was heated up to 60°C and mixed with 5 bars of *para*-hydrogen. Subsequently a polarization sequence introduced by Goldman *et al.* was applied.[S3] After a 20 s transport the NMR signal was detected in a 9.4 T 94/20 Bruker Biospec. This experiment was performed twice after optimization of the pulse sequence. In order to determine the polarization, the hyperpolarized signal was compared to its signal in thermal equilibrium averaged 200 times. *J*-couplings and parameters applied for the sequence are as follows (refer to scheme S1 for the derivative) and a schematic drawing of the sequence can be found in figure S9:

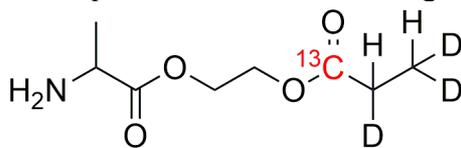

**Scheme S1.** Alanine-derivative that was polarized at the indicated $^{13}$C nuclei using a polarization transfer sequence.

*J*-couplings: $^3J_{H,H}$ = 7.24 Hz; $^3J_{13C,H}$ = -5.62 Hz; $^2J_{13C,H}$ = 7.57 Hz

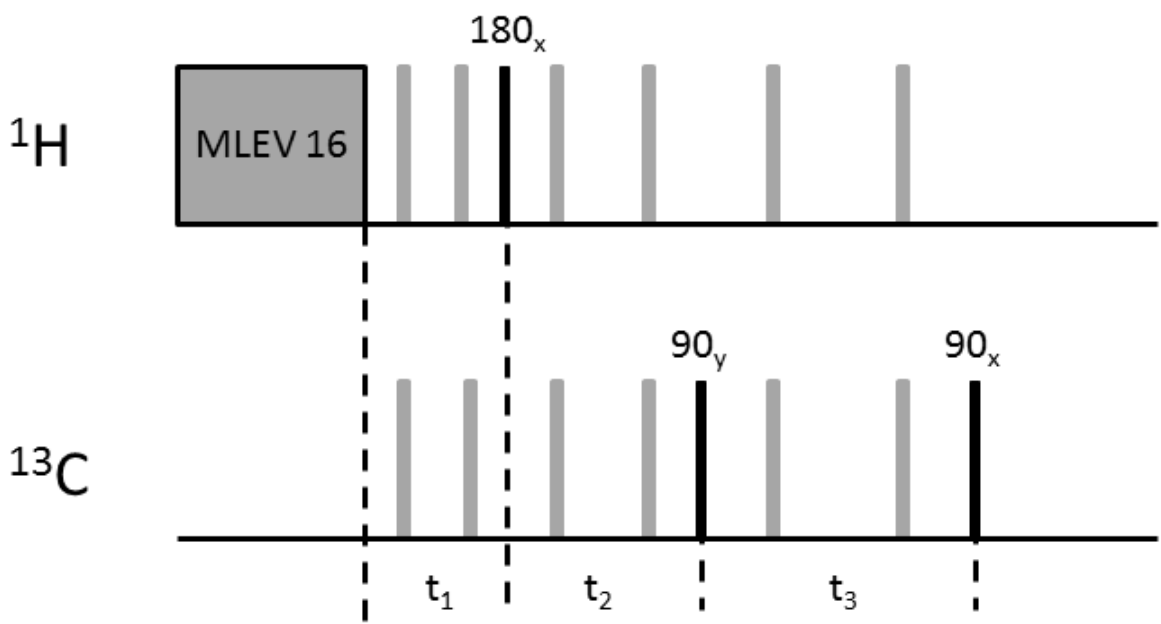

**Fig. S9.** Schematic of the Goldman sequence used. Grey bars indicate 180° pulses at one fourth and three forth of the evolution periods. The black bars are the pulses important for the polarization transfer.[S5]

The timings are: $t_1$ = 28.28 ms , $t_2$ = 36.20 ms, $t_3$ = 50.34 ms

## S5 Relaxation times of the investigated molecules

The longitudinal relaxation times ($T_1$) of the synthesized molecules were investigated after hydrogenation with an AV500 system ($B_0$ = 11.7 T) with cryoprobe for the protonated derivatives with an inversion recovery experiment and one scan. For carbon experiments 8 scans were used with an inversion recovery experiment with inverse gated decoupling . The $T_1$ for the deuterated and $^{13}$C labelled derivative was recorded with an AV600 system ($B_0$ = 14.1 T) with 1 scan for protons and 8 scans for carbon. All experiments were performed at room temperature. HEP, Ala-HEP and Ac-Ala-HEP were investigated in D$_2$O in the presence of air whereas for solubility and stability reasons $T_1$ of the N-protected Gly-HEP and Gln-HEP derivatives were measured in methanol-d$_4$. The pH of the samples was 6.5±0.5 and the catalyst concentration 10 mol% of the material, which means 2.5 mM.

HEP:

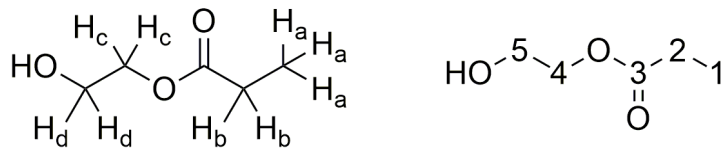

**Table S1. Longitudinal $^1$H relaxation times of HEP**

| $^1$H position | a | b | c | d |
|---|---|---|---|---|
| $T_1$/s | 4.0 | 3.5 | 2.3 | 2.1 |

**Table S2. Longitudinal $^{13}$C relaxation times of HEP**

| $^{13}$C position | 1 | 2 | 3 | 4 | 5 |
|---|---|---|---|---|---|
| $T_1$/s | 5.5 | 4.6 | 15.9 | 1.9 | 2.0 |

Ala-HEP:

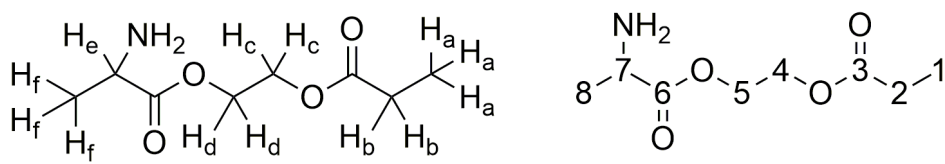

**Table S3. Longitudinal $^1$H relaxation times of Ala-HEP**

| $^1$H position | a | b | c | d | e | f |
|---|---|---|---|---|---|---|
| $T_1$/s | 4.1 | 3.0 | 0.9 | 0.9 | 5.9 | 1.2 |

**Table S4. Longitudinal $^{13}$C relaxation times of Ala-HEP**

| $^{13}$C position | 1 | 2 | 3 | 4 | 5 | 6 | 7 | 8 |
|---|---|---|---|---|---|---|---|---|
| $T_1$/s | 5.9 | 5.8 | 14.2* | 1.0 | 1.1 | 10.2 | 2.4 | 1.3 |

*T1 for the deuterated derivative corresponds to 21.5s

Ac-Ala-HEP:

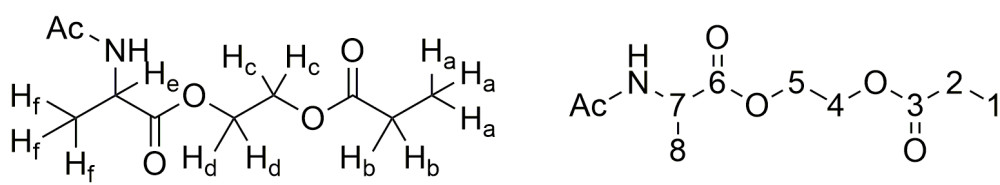

**Table S5. Longitudinal $^1$H relaxation times of Ac-Ala-HEP**

| $^1$H position | a | b | c | d | e | f |
|---|---|---|---|---|---|---|
| $T_1$/s | 3.4 | 3.2 | 2.4* | 1.0 | 2.4* | 0.7 |

*chemical shift values of c and d are merged. Therefore the values reported reflect a time extracted from a monoexponential fit over the two proton signals

**Table S6. Longitudinal $^{13}$C relaxation times of Ac-Ala-HEP**

| $^{13}$C position | 1 | 2 | 3 | 4 | 5 | 6 | 7 | 8 |
|---|---|---|---|---|---|---|---|---|
| $T_1$/s | 4.8 | 4.3 | 15.6 | 0.8 | 0.9 | 4.2 | 4.3 | 1.1 |

Boc-Gly-HEP:

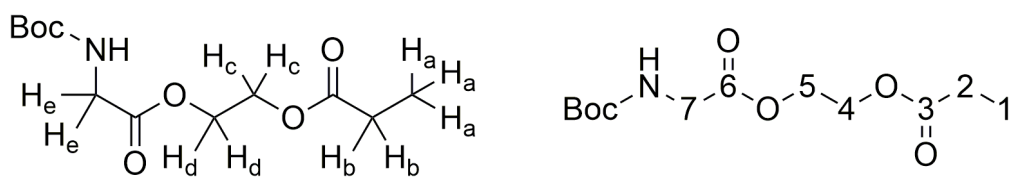

**Table S7. Longitudinal $^1$H relaxation times of Boc-Gly-HEP**

| $^1$H position | a | b | c | d | e |
|---|---|---|---|---|---|
| $T_1$/s | 5.5 | 5.1 | 2.0 | 1.7 | 1.7 |

## Table S8. Longitudinal $^{13}$C relaxation times of Boc-Gly-HEP

| $^{13}$C position | 1 | 2 | 3 | 4 | 5 | 6 | 7 |
|---|---|---|---|---|---|---|---|
| $T_1$/s | 6.3 | 6.1 | 16.7 | 1.6 | 1.5 | 11.1 | 1.8 |

Boc-Gln(Xan)-HEP:

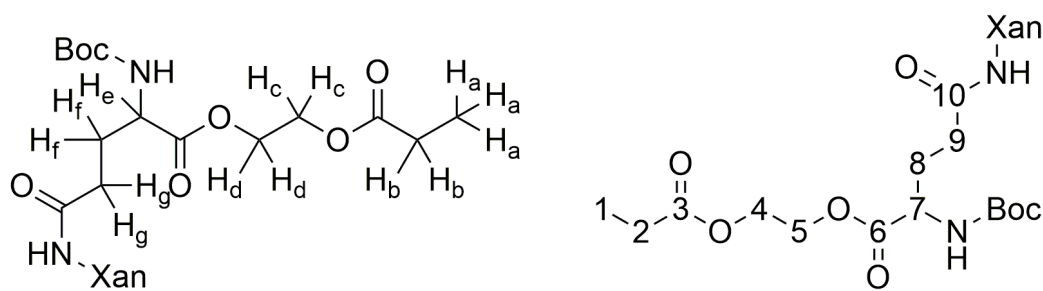

## Table S9. Longitudinal $^1$H relaxation times of Boc-Gln(Xan)-HEP

| $^1$H position | a | b | c | d | e | f | g |
|---|---|---|---|---|---|---|---|
| $T_1$/s | 3.6 | 2.8 | 1.2 | 1.2 | 2.0 | 4.3 | 4.3 |

## Table S10. Longitudinal $^{13}$C relaxation times of Boc-Gln(Xan)-HEP

| $^{13}$C position | 1 | 2 | 3 | 4 | 5 | 6 | 7 | 8 | 9 | 10 |
|---|---|---|---|---|---|---|---|---|---|---|
| $T_1$/s | 6.6 | 4.8 | 15.6 | 0.4 | 0.4 | 7.9 | 3.3 | 1.8 | 2.9 | 11.1 |

# S6 Catalyst decomposition

When an NMR spectrum is recorded as part of a PHIP experiment in which the reaction mixture was heated up to 80 °C, additional hyperpolarized peaks could be observed at 0.9, 1.8, 3.1 and 6.2 ppm. Norbornadiene is part of the used metal complex that catalyzes the hydrogenation of the amino acid derivatives. The observed additional peaks correspond to norbornene indicating that at elevated temperature the norbornadiene is partly hydrogenated as the catalyst decomposes.[S4,S5] Figure S10 shows a $^1$H NMR spectrum of the hyperpolarized Ala-HEP derivative and hyperpolarized norbornene.

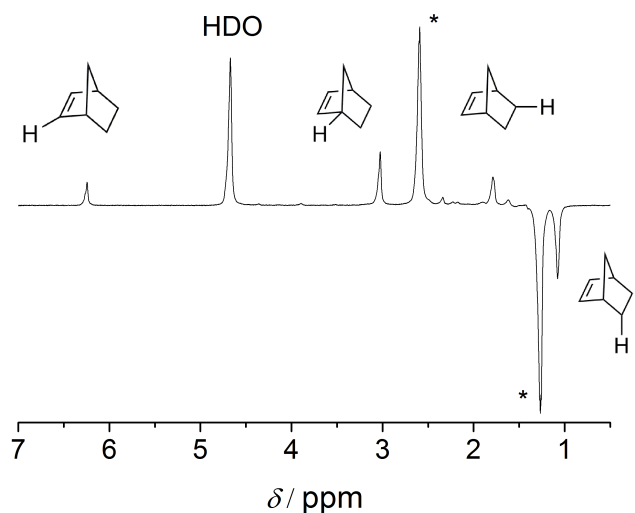

**Fig. S10.** $^1$H NMR spectrum of hyperpolarized norbornene and Ala-HEP at $B_0 = 14.1$ T (600 MHz $^1$H frequency) after a 45° pulse. Asterisks indicate the propionate part of the hyperpolarized amino acid derivative.

## S7 Remarks on previous amino acid PHIP

PHIP studies on amino acids (non-derivatives) in water have been demonstrated as a proof of principle, but the achieved polarization showed only small improvements over thermal equilibrium conditions, which prohibited any useful *in vivo* applications.[S6] Currently, the most potent r.f. polarization transfer sequence from $^1$H to $^{13}$C was developed by Goldman and coworkers.[S3] Except for the previously polarized γ-aminobutyric acid (GABA), it is unlikely that this sequence can be used to generate high polarization levels in the carbonyl $^{13}$C (typically a long-lived $^{13}$C species) of the reported unprotected amino acids. This is due to the small *J*-couplings (in particular of $^4J$ of aliphatic protons to $^{13}$C) leading to unfavorable conditions (e.g., timing and angle) with regards to transfer of proton polarization (see above for pulse sequence details or ref [S3]). γ-aminobutyric acid has the right structure for efficient polarization transfer from protons to the desired long-lived carbonyl $^{13}$C. But the chemical reaction appears to be too slow to efficiently transfer spin order of p-H$_2$ and it is necessary to protect the catalyst by working in a very acidic environment to achieve any polarization at all.[S3] Working in an acidic environment, in the context of potential *in vivo* applications, means that the polarized substrate needs to be diluted in a buffer as an additional step before injection, which may be a source of polarization loss. No reported research has achieved a $^{13}$C PHIP polarization of N-unprotect amino acids on the order of 1%, a prerequisite for in vivo applications.

# S8 Toxicity and Metabolic considerations of hydrolysis

Previous PHIP studies utilizing esters revealed that esters are quickly taken up in cells and are hydrolyzed fast by enzymatic processes.[S7] As the synthesized amino acid derivatives are based on esters we would like to provide some considerations regarding the time of the hydrolysis, potential metabolites and their potential toxicity.

The synthesized derivatives contain two ester moieties that may potentially be hydrolyzed. It is plausible that the following chemicals could be observed due to enzymatic hydrolysis: free amino acid, ethylene glycol and propionic acid. Ethylene glycol and propionic acid originate from HEP. The amino acids used in this study can be seen as non-toxic as glycine, glutamine and alanine are metabolites in the human body. Ethylene glycol has been identified to have toxic effects on humans. Studies have revealed that a threshold for ethylene glycol to have toxic effects on humans is a concentration of about 3.2 mM in the plasma.[S8] In typical *in vivo* experiments utilizing hyperpolarized contrast agents, concentrations of 25-300 mM in 1 mL are administered.[S9] Considering that humans carry multiple liters of blood plasma, the concentration of formed ethylene glycol in a potential *in vivo* experiment is more than one order of magnitude lower than the described threshold and thus the expected toxicity should be low. Propionic acid is a metabolite that is transformed into succinyl-CoA, which is an intermediate of the citric acid cycle.[S10] Therefore, propionic can be seen as a substrate with very low toxicity and can also be seen as a potential new tracer for the citric acid cycle.

Regarding the speed of hydrolysis, we estimate it to be on a one to several minutes timescale, which would be sufficient for the proposed *in vivo* experiments. Experiments with esters of fatty acids have shown that a half-life of 58 s can be expected for simple esters derived from ethanol.[S11] With the more sterically demanding properties of the synthesized esters, the hydrolysis may take longer as has also been observed in the context of prodrug research making use of ester moieties.[S12] Although parts of the potential contrast agents will be hydrolyzed, we expect that sufficient polarization can be delivered *in vivo* in the expected time scale.